# A method for energy estimation and mass composition determination of primary Cosmic rays at Chacaltaya observation level based on atmospheric Cerenkov light technique

*S. Cht. Mavrodiev*[1], *A. Mishev*[1,2,*], *J. Stamenov*[1]

[1]INRNE-BAS, 72 Tzarigradsko chose, 1784 Sofia, Bulgaria

[2]LIP av. Elias Garcia 14-1, 1000-149 Lisboa Portugal

* Corresponding author: LIP, av. Elias Garcia 14-1, 1000-149 Lisboa PortugalTel (+351) 217 973 880; fax: (+351) 217 934 631; e-mail: aleksandar.mishev@cern.ch

**Abstract:** A new method for energy and mass composition estimation of primary cosmic ray radiation based on atmospheric Cerenkov light flux in extensive air showers (EAS) analysis is proposed. The Cerenkov light flux in EAS initiated by primary protons and iron nuclei is simulated with CORSIKA 5.62 code for Chacaltaya observation level (536 g/cm$^2$) in the energy range 10 TeV – 10 PeV. An adequate model, approximation of lateral distribution of Cerenkov light in showers is obtained. Using the proposed model and solution of overdetermined system of nonlinear equations based on Gauss Newton method with autoregularization, two different array detector arrangements are compared. The detector response for the detector sets is simulated. The accuracies in energy and shower axis determination are studied and the corresponding selection criteria are proposed. An approximation with nonlinear fit is obtained and the energy dependence of the proposed model function parameters is studied. The approximation of model parameters as function of the primary energy is carried out. This permits, taking into account the properties of the proposed method and model, to distinguish proton primaries from iron primaries. The detector response for the detector sets is simulated and the accuracies in energy determination are calculated. Moreover the accuracies in shower axis determination are studied and criteria in shower axis position estimation are proposed.

## 1. Introduction

In cosmic ray physics are many interesting problems such as their origin, acceleration and interactions. One of the principle problems in the field of primary cosmic ray investigations is the energy spectrum and mass composition precise determination. This is very important in order to obtain some information about the cosmic ray origin, propagation mechanisms and interstellar matter. In the region of ultra high energies only indirect measurements are possible. The estimation of the energy and the nature of the primaries based on ground observations is very difficult, because the high level of the shower development and the experimental noises, connected with EAS. The development of independent or not used previously techniques can give us more effective methods, for example, on the basis of inverse problem solutions.

Other very important aim is the statistic amelioration. One of the possible techniques of investigation is based on atmospheric Cerenkov light measurements in EAS, precisely obtaining the lateral distribution function.

At Chacaltaya observation level 536 g/cm$^2$ the possibility to determine the mass composition of primary cosmic ray radiation using different lateral distributions of EAS components, their fluctuations at different distances seems acceptable [1]. So the decision to use atmospheric Cerenkov technique is taken in attempt to check some new possibilities for primary cosmic ray investigation.

## 2. The method

In the general case the lateral distribution function of Cerenkov light in EAS depends on the energy E and the type of the initiating primary particle, the distance R from the shower axis, the observation level the height of first interaction etc…

$$Q = Q(R, E, \theta, \varphi, H, H_0, \alpha) \tag{1}$$

where R is the distance from the shower axis, E the energy of the initiated primary particle, $\theta, \varphi$ are the astronomical coordinates of the shower (zenithal and azimuthall angles), H the observation level, $H_0$ is the level of the shower birth (the beginning of the cascade) and $\alpha$ is a parameter, which depend of the type of the primary particle.

For a concrete experiment (for exemple the HECRE proposal at Chacaltaya) [2] the observation level is given. In the case of simulated events the level of first interaction is part of the physical fluctuations of the processes and per consequence it is included in the model. In the case of vertical events the dependece of the zenithal angle not exist nor the azimuthal. Finally the lateral distribution of Cerenkov light densities is function of R, the distance from the shower axis and a few parameters p, which are function of the type and thy energy of the primary particle in terms above

$$Q = Q[R, p(E, \alpha)] \tag{2}$$

With CORSIKA [3] code using the VENUS [4] and GHEISHA [5] like hadronic models at Chacaltaya observation level 536 g/cm$^2$ in the energy range 10 Tev –10 PeV for proton and iron primaries few charateristiques of EAS are simulated, precisely the Cerenkov light flux densities, muon, hadronic and electromagnetic component. One large detector 800x800m is used to obtain the lateral distributiuon of Cerenkov light. The aim is to reduce the statistical fluctuations.

There are few reasons to choose this observation level. First of all the existing experimental proposal HECRE [2]. This observation level is near to the shower maximum and the fluctuations in the shower development are not so important and it is possible to obtain more flat functions.

For each energy 500 EAS are simulated. The obtained lateral distributions of Cerenkov light densities are presented in fig.1 (dote lines). Near to the shower axis the lateral distribution has more or less plate maximum and after several hundred meters (depending on the primary and its energy) it is exponentially decreasing to zero. The model function of the atmospheric Cherenkov light distribution has to be in a class of functions with such behavior. Different criteria are used to search the model function. It's clear that the model must be a good fit. For the proposed method is very important that the behavior of model parameters in function of the energy is monotone. Finally the model must be an integrable function [6, 7].

Using REGN [8] code and solving inverse problem - overdetermined system of equations the approximation based on Gauss Newton method with autoregularization is carried out fig.1 (solid lines). The proposed approximation is in the class of Breit Wigner function.

$$Q(R) = \frac{\sigma e^{a} e^{-\left[\frac{R}{\gamma}+\frac{R-r_0}{\gamma}+\left(\frac{R}{\gamma}\right)^2+\left(\frac{R-r_0}{\gamma}\right)^2\right]}}{\gamma\left[\left(\frac{R}{\gamma}\right)^2+\left(\frac{R-r_0}{\gamma}\right)^2+\frac{R\sigma^2}{\gamma}\right]} \qquad (3)$$

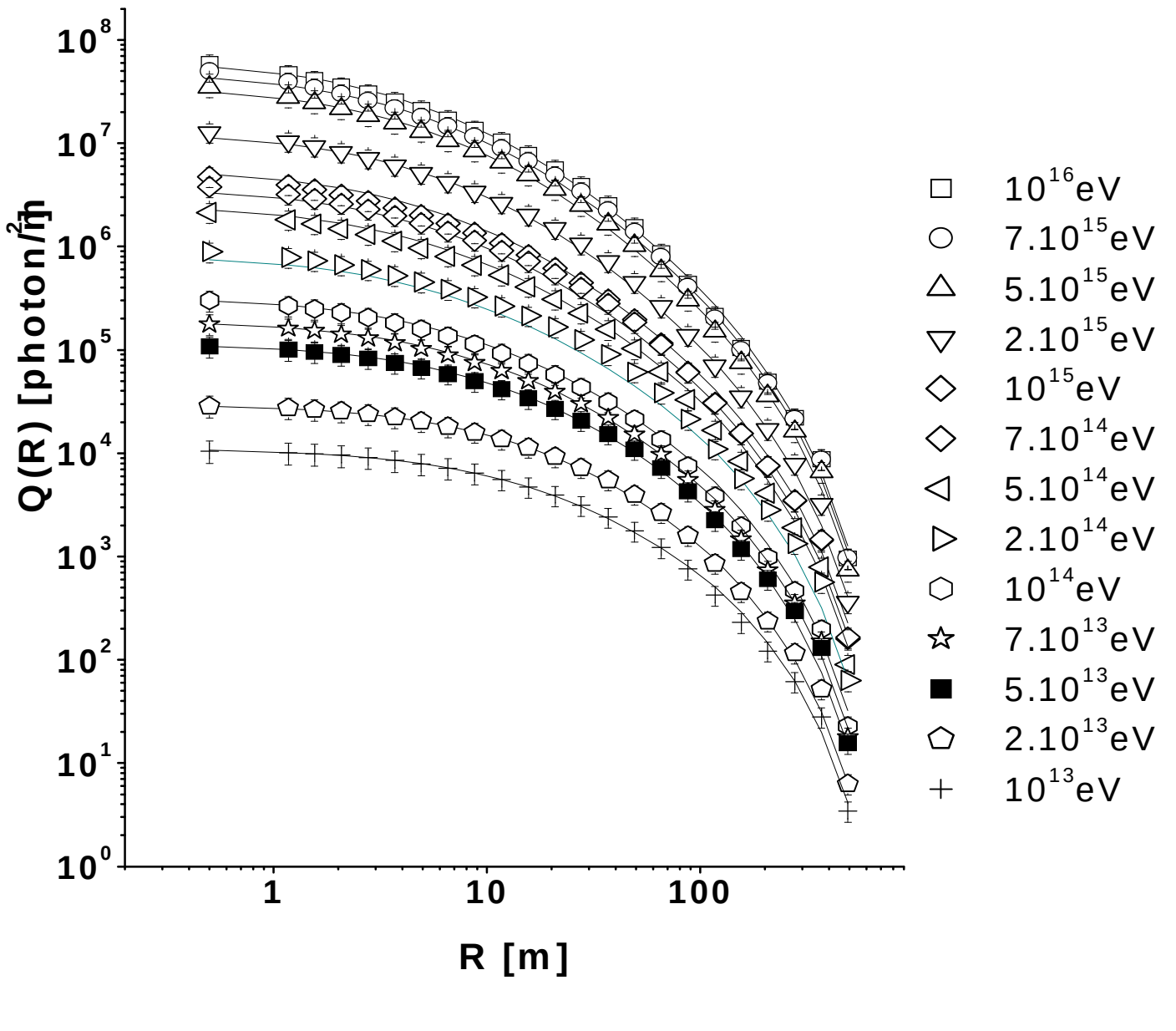

Fig.1 Lateral distribution of Cerenkov light (dote line) simulated with Corsika code and the obtained approximation (solid line) for primary protons

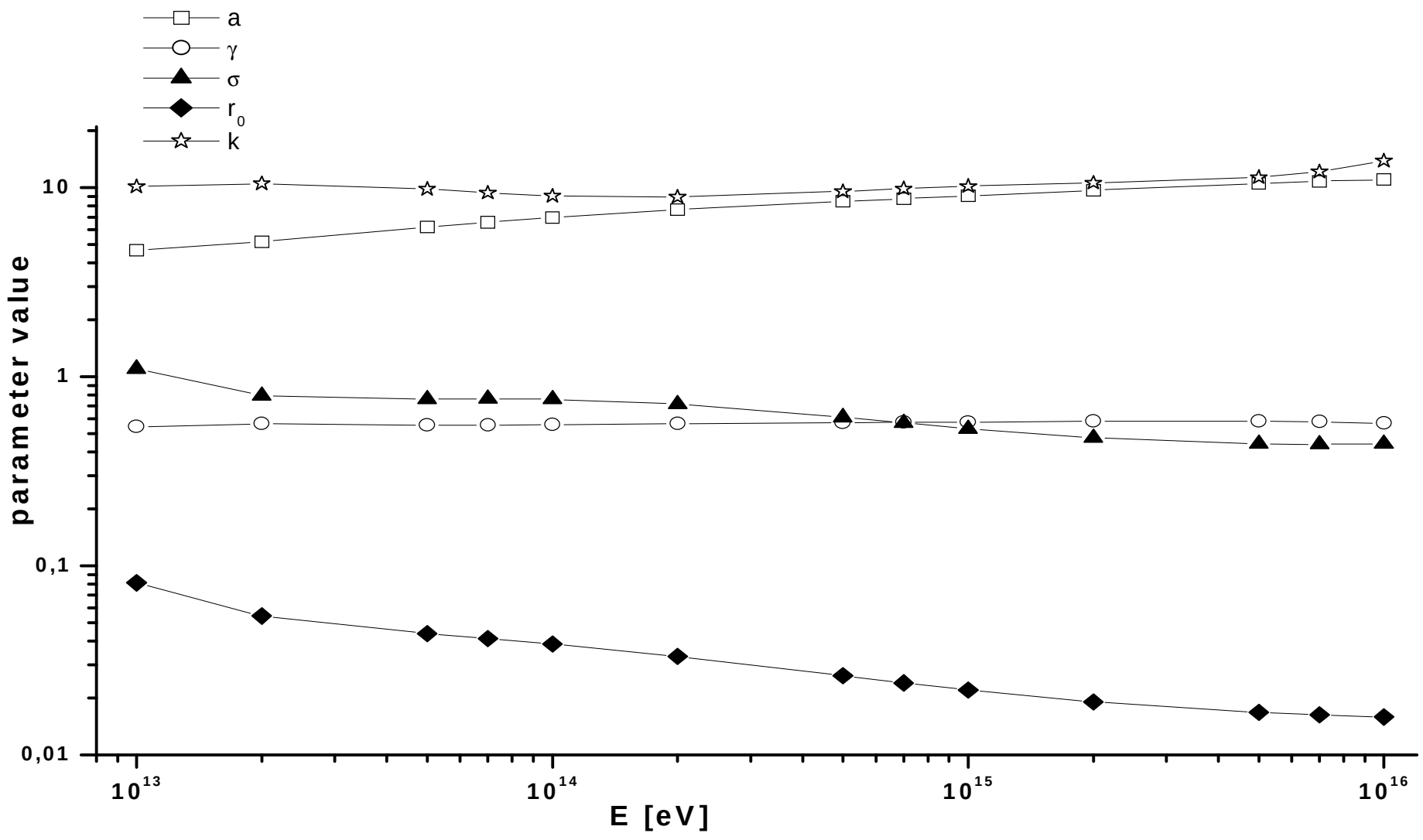

Fig.2 The model parameters values for primary protons in function of the energy

where R is the distance from the shower axis, and $\sigma$, $\gamma$, a and $r_0$ are the model parameters. In the model is used that the energy of the incident primary particle is equal to the total number of Cerenkov photons at the given observation level multiplied by a parameter k.

$$E = \kappa f(N_q) \quad (4)$$

where $N_q$ is the total number of Cerenkov light photons

$$N_q = 2\pi \int Q(R) dR \quad (5)$$

This is the main reason to search an integrable function for the model. The behavior of model parameters in function on the energy is presented in fig.2.

After that the model parameters are also approximated and replaced with their approximations in the initial model. The difference between the initial fit and final is less then 2%. The result is that finally we have a model only with two variables R and E.

Similar calculations are carried out for iron primary particle. The lateral distributions for different incident particles are different (see fig. 3), nevertheless the same model function (see eq. 3) is used for the approximation, because the different values of function parameters for different primaries. The results are presented in fig. 4 (dote lines are the simulated with Corsika lateral distributions and the solid lines are the obtained approximation). The difference between proton and iron in model parameter behavior in function on the energy is presented in fig.5. The strong nonlinearity of the model and this difference permits to distinguish the initiating primaries (proton from iron) on the basis of the different $\chi^2$. For example the $\chi^2$ for protons became 10 times larger using the parameterization of the iron fit.

So one can summarize: for simulated with Corsika code data at Chacaltaya observation level 536 g/cm$^2$ a model approximation of the lateral distribution function of Cerenkov light in EAS is obtained for distances up to 450 m from the shower axis. This model is obtained for proton and iron primaries in the very interesting energy range- the region around the knee.

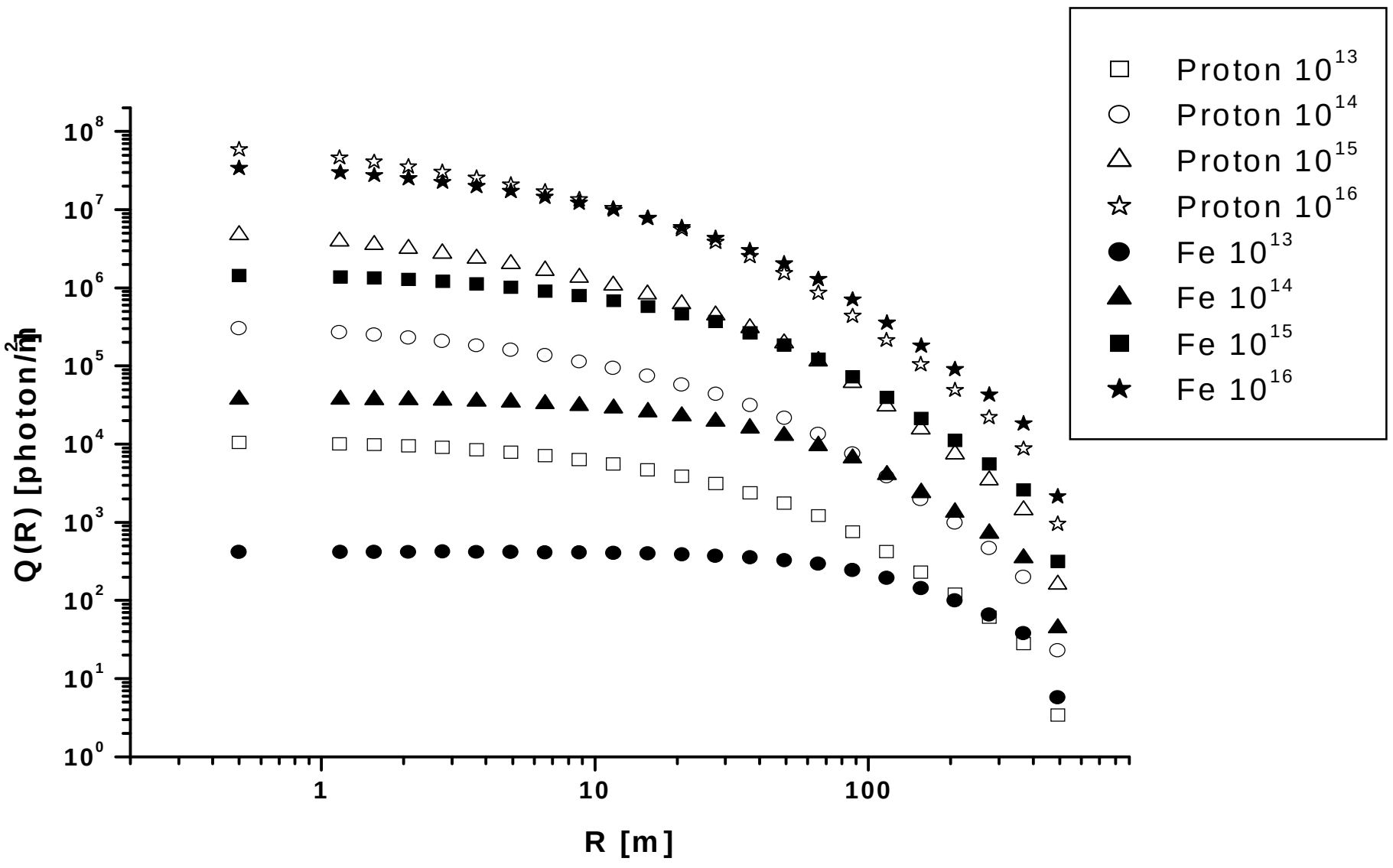

Fig.3 Difference between primary proton and iron lateral distributions of Cerenkov light

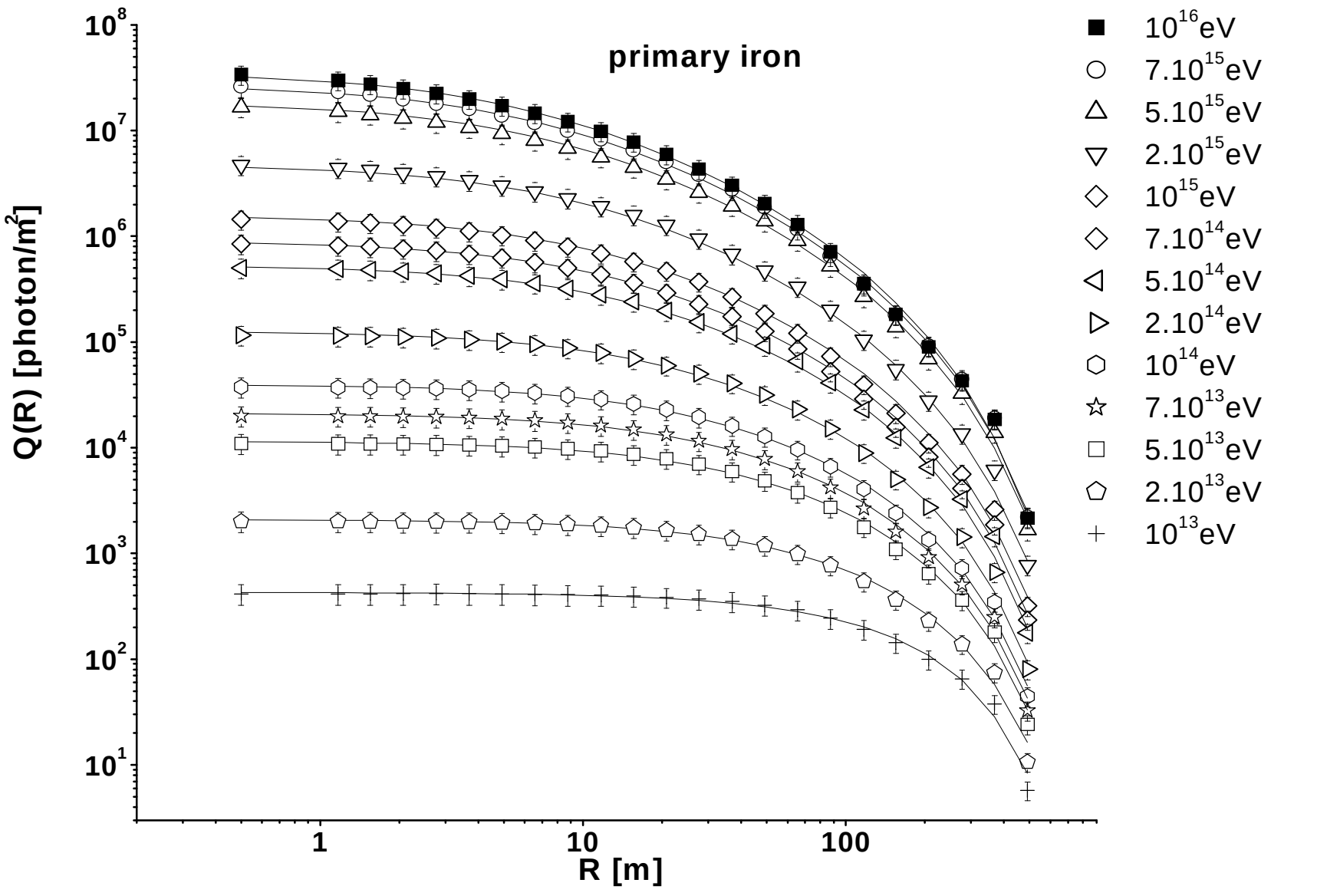

Fig.4 Lateral distribution of Cerenkov light (dote line) simulated with Corsika code and the obtained approximation (solid line) for primary iron

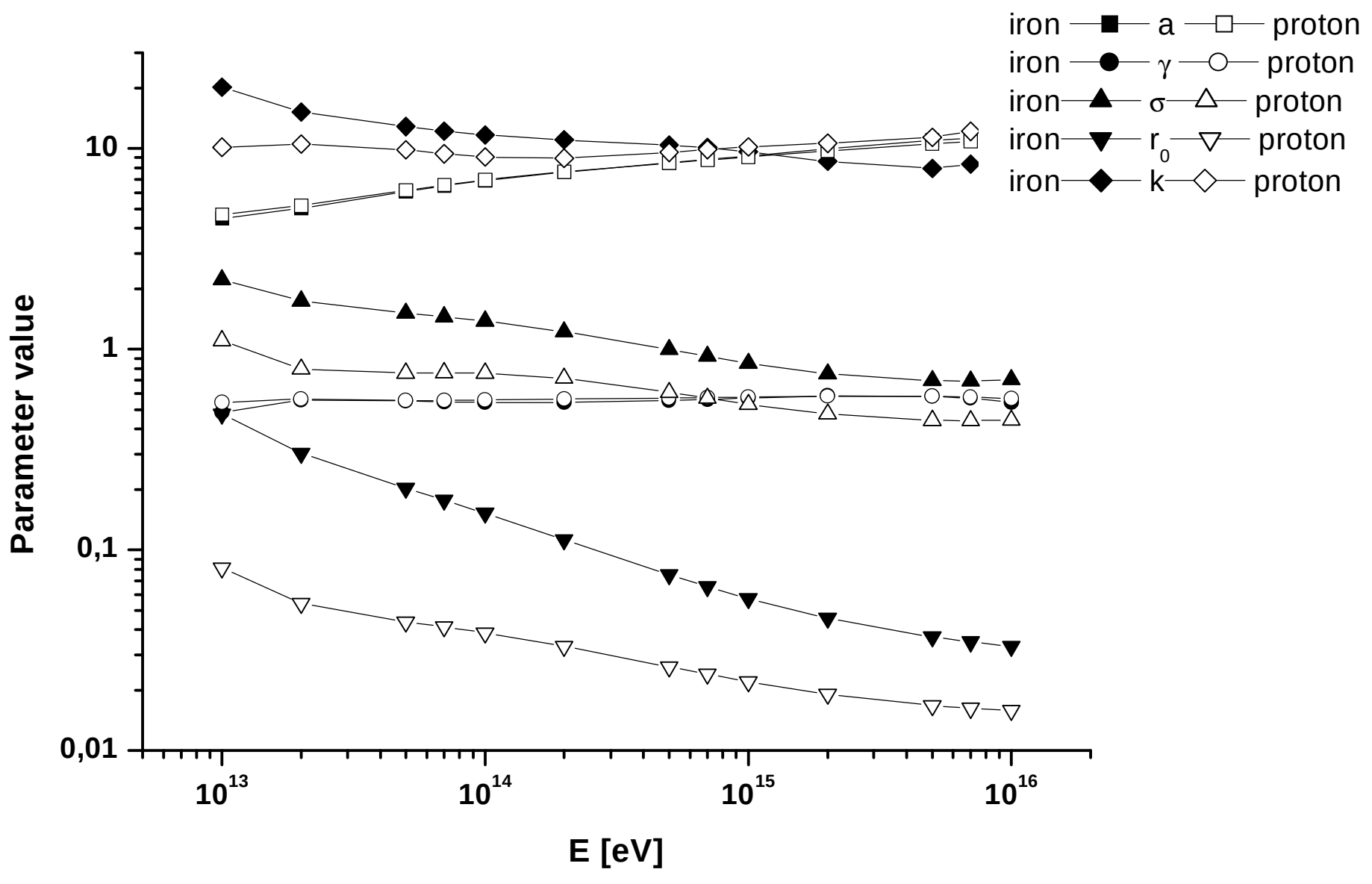

Fig.5 Difference between proton and iron for the model parameters behavior in function of the energy

## 3. Detector response simulation

The proposed model gives the possibility to simulate the detector response of atmospheric Cerenkov detectors and calculate the accuracies in energy estimation using quasiexperimental data. The first one is according HECRE proposal fig.6 . This is a uniform set of 49 detectors like AEROBICC [9].

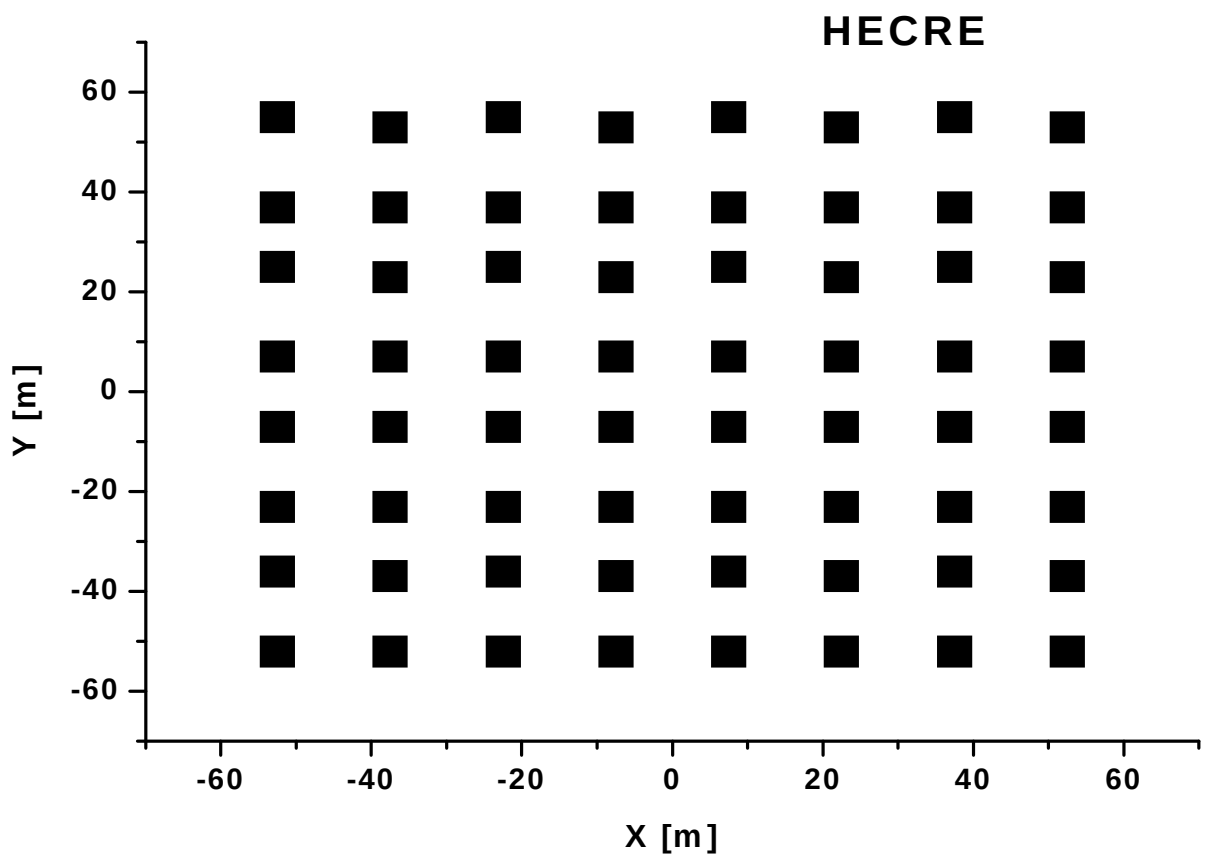

Fig.6 HECRE detector array

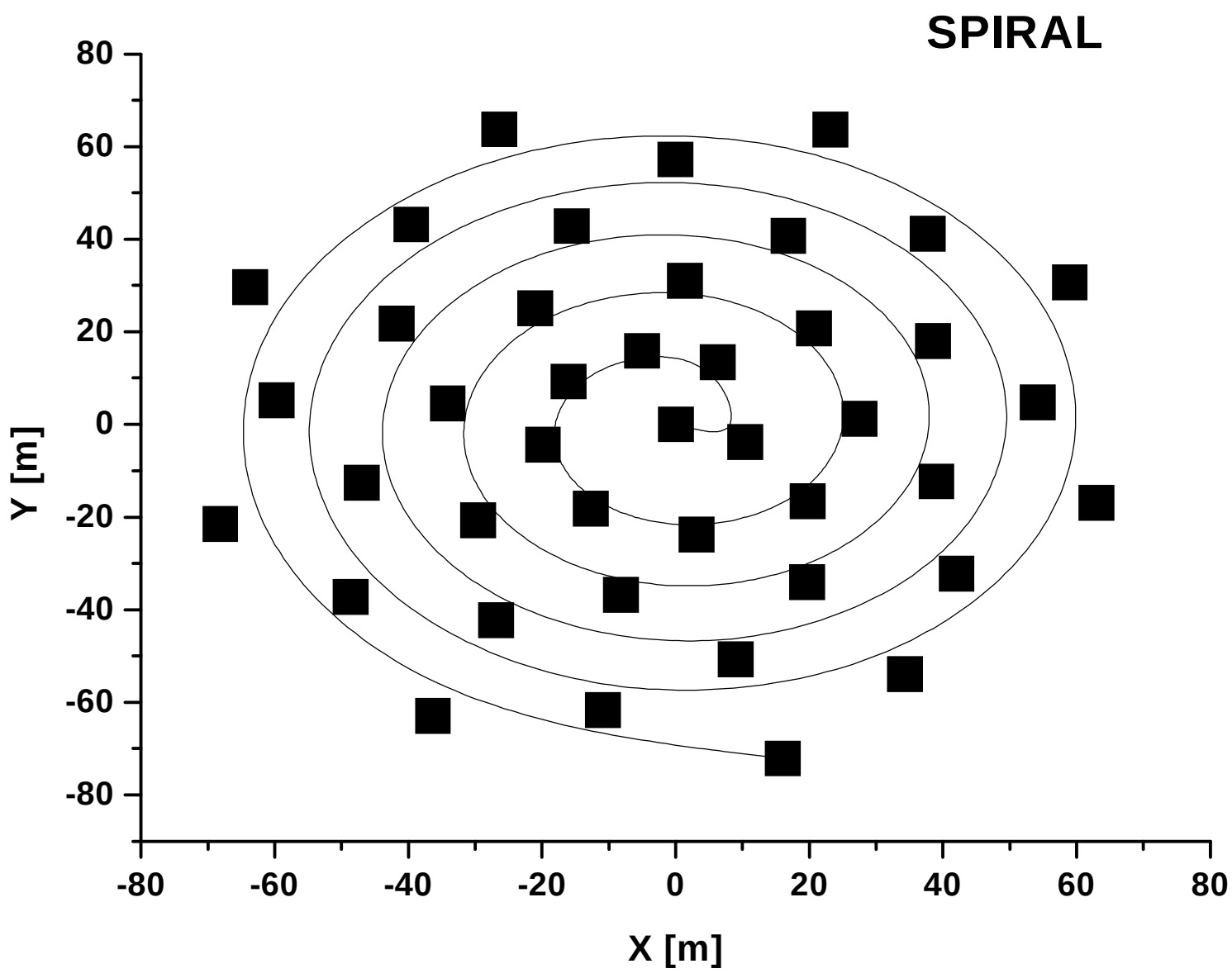

Fig.7 SPIRAL detector array

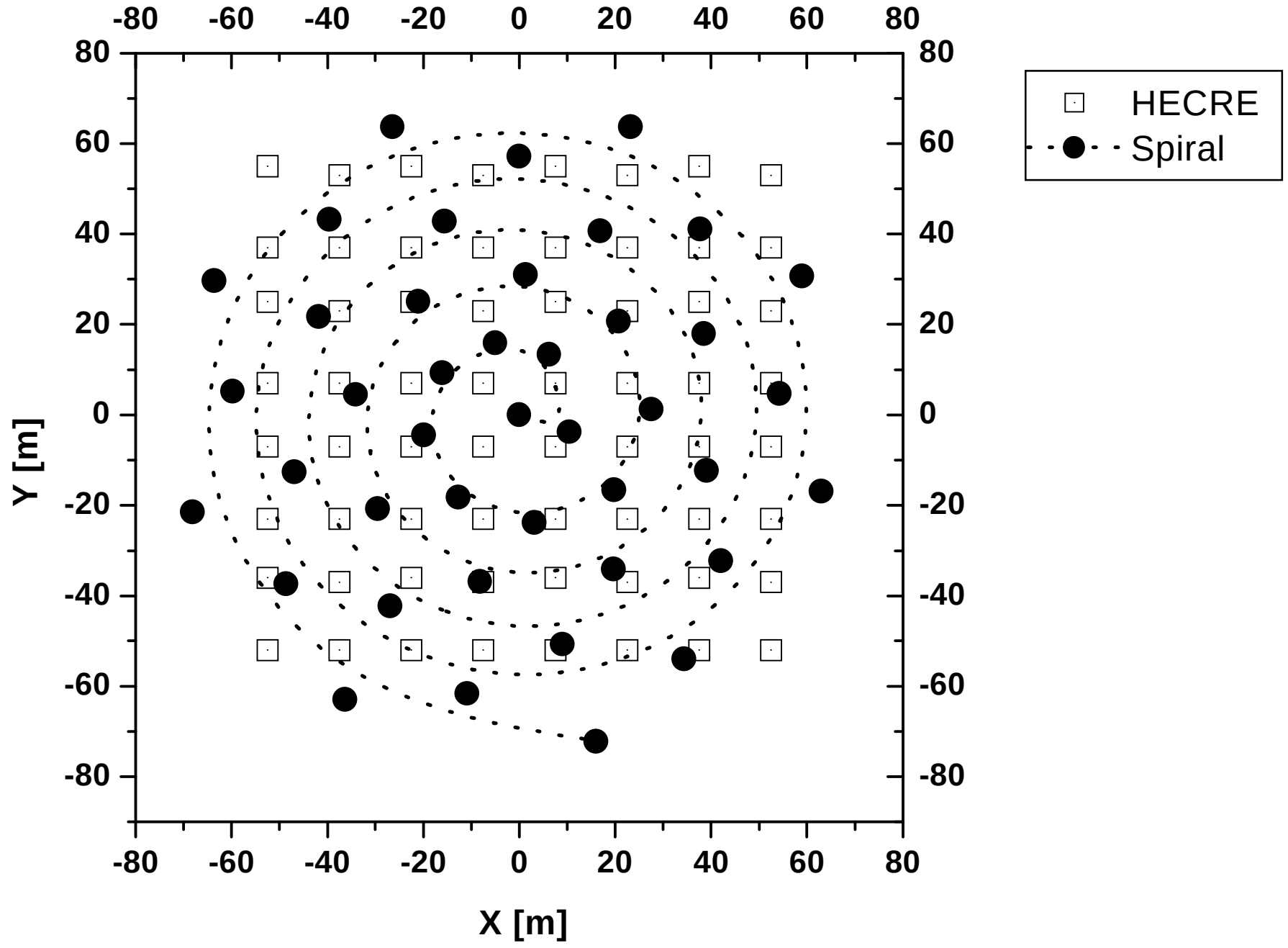

Fig.8 Difference between HECRE and SPIRAL

The obtained approximation function gives the number of Cerenkov photons per $m^2$ and it is possible to calculate the number of Cernkov photons in detector taking into account the detector surface. A proper code is developed for simulation the detector response. The energy of the shower is simulated according step spectrum, an uniform azimuthall angle distribution and uniform distribution of shower axis in the field of detectors. After that this number of Cerenkov photons in the detector is recalculated, according Poisson or Gauss distribution depending of quantity of Cerenkov photons in the detector.

The first simulation is made out for 5000 events primary protons in very interesting energy range around the knee. One more time an inverse problem solution is carried out with the simulated events in two cases 30% and 50% additional systematic error. This error summarize all systematic and registration errors of the device. Similar investigations are made for second one type of detector displacement spiral detector (logarithmic spiral without few point near the center) fig.7. The difference between two detector sets is show in fig.8.

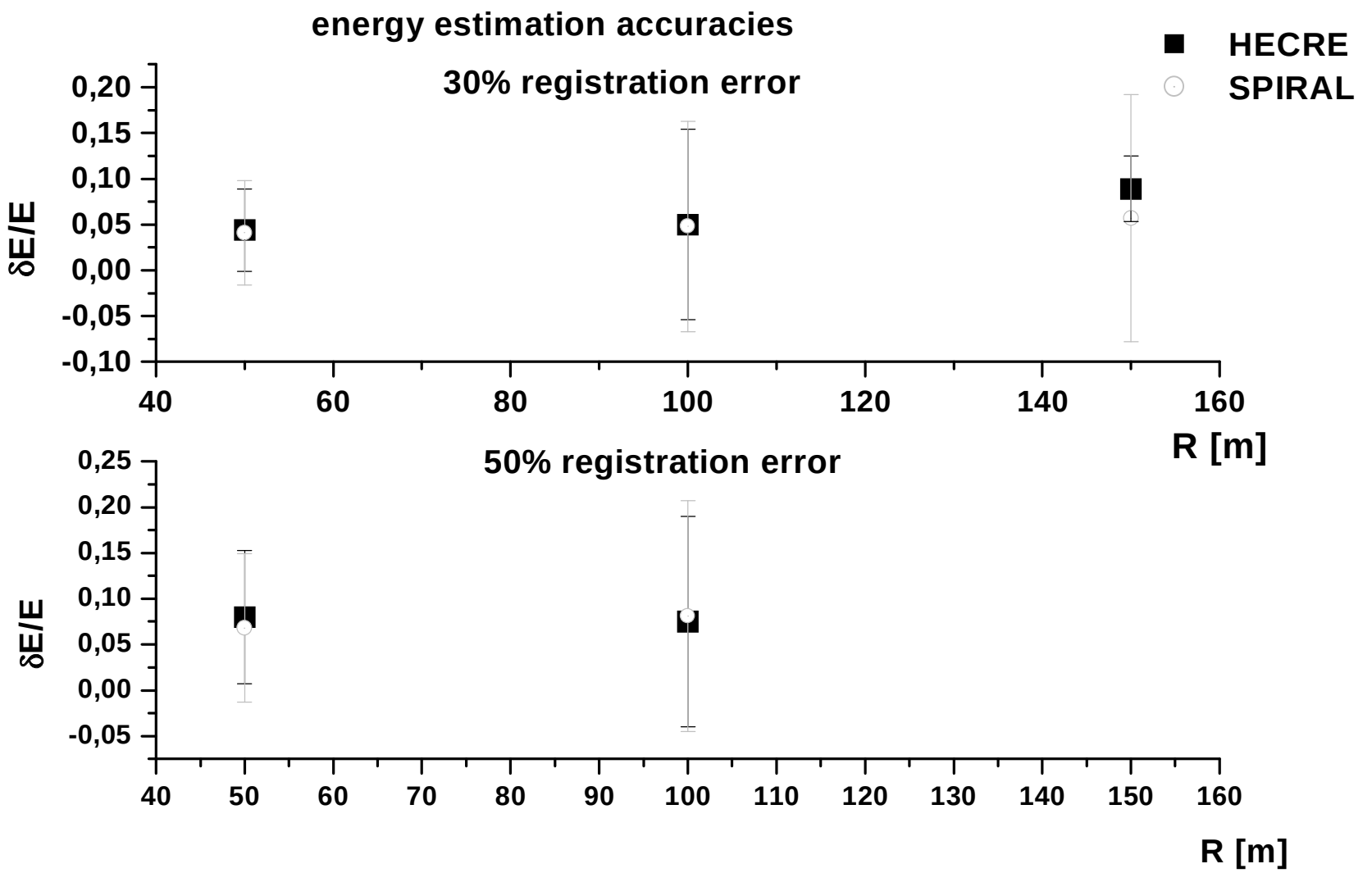

Fig.9 Energy estimation accuracies for HECRE and SPIRAL for different shower axis from the center of detector array

The number of solute events in both cases are similar nevertheless the less number of detectors in SPIRAL. In fig. 9 are presented the obtained accuracies in energy estimation in the case of 30% and 50% registration error and shower axis at distances from center of the array 50, 100 and 150m. The obtained accuracies for energy estimation are less then 15%.

Moreover, an additional simulation with simplified mixed mass composition containing 70% protons and 30% iron is carried out. This permits additionally to check the mass composition determination and energy estimation simultaneously. In this case the relative number of solute events diminishes with only 5 %, compared with similar results for energy estimation accuracy estimation.

The case when the obtained $\chi^2$ is big according the method criteria and it is impossible to give any information about the primary there are showers with big fluctuations or big registration error and the shower axis are more then 100 m from the center of detector array.

Using the obtained $\chi^2$ we propose criteria, which permit to estimate with adequate precision the shower axis (fig.10) and solve the problem of energy estimation of showers with axis far away from the detector center.

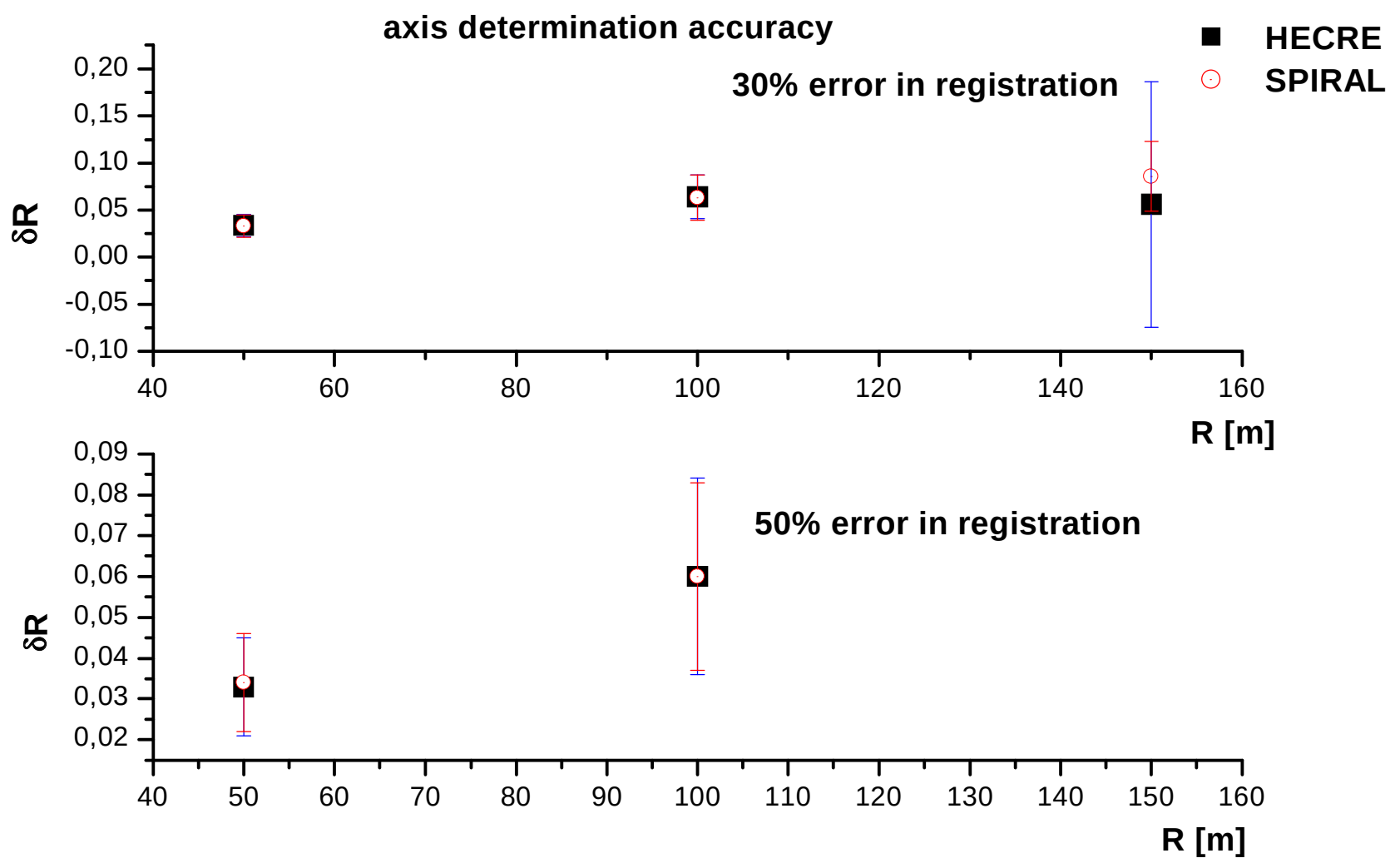

Fig.10 Shower axis determination accuracies for HECRE and SPIRAL for different shower axis from the center of detector array

**Conclusion**

The lateral distributions of atmospheric Cerenkov Light in EAS initiated by protons or nuclei ions nuclei with energies 10 TeV- 10 PeV were obtained with help of the Corsica code [3], using VENUS [4] and GHEISHA [5] hadronic models for Chacaltaya observation level 536 g /sm$^2$. The lateral distributions were approximated with nonlinear function, which analytical form and parameters values were obtained by solving overedetermined systems using REGN code [6]. The obtained solutions permit to estimate the energy and the nature of the initiating primary particle.

Additionally are analyzed two possible detector displacement of the corresponding EAS array. It is shown, that the uncertainties by the primary energy estimations not exceed 15 – 20 % and the reached shower axis coordinates accuracy is sufficient to apply the proposed new EAS selection parameters based only on Cerenkov light measurements in EAS [10]. Moreover it is shown, that the proposed method gives the possibility to separate EAS initiate by primary protons and ions nuclei in quasi real time using only the information from 30 detectors displaced according a spiral set to 150 m from the array centre at observation level 536 g /sm$^2$.

In future deeper observation levels will be analyzed with the same method in attempt to study its applicability for other EAS components for researched the energy distribution and chemistry of cosmic ray.

**Acknowledgements**

We are thank full of IT division at INRNE for the given computational time and assistance.